\documentclass[twocolumn,aps]{revtex4}
\usepackage{amsfonts}
\usepackage{amsmath}
\usepackage{bm}
\usepackage{graphicx}
\usepackage{xcolor}
\usepackage{color}
\usepackage{ amssymb }

\def \beq {\begin{equation}}
\def \eeq {\end{equation}}
\def \bea {\begin{eqnarray}}
\def \eea {\end{eqnarray}}
\def \bfig {\begin{figure}}
\def \efig {\end{figure}}

\def \lab {\label}

\def \nn {\nonumber}
\def \de {\partial}

\def \fr {\frac}

\begin{document}

\title{Magnetosperic Multiscale (MMS) 
observation of plasma  velocity-space cascade: Hermite representation and theory}

\author{ S. Servidio$^1$, A. Chasapis$^2$, W. H. Matthaeus$^2$, D. Perrone$^3$, F. Valentini$^1$, T. N. Parashar$^2$,  P. Veltri$^1$, D. Gershman$^4$, C. T. Russell$^5$, B. Giles$^4$, S. A. Fuselier$^6$, T. D. Phan$^7$, J. Burch$^5$ }

\affiliation{
$^1$Dipartimento di Fisica, Universit\`a della Calabria, I-87036 Cosenza, Italy\\
$^2$Bartol Research Institute and Department of Physics and Astronomy, University of Delaware, Newark, DE 19716, USA\\
$^3$European Space Agency, Science and Robotic Exploration Directorate, ESAC, Madrid, Spain\\
$^4$NASA Goddard Space Flight Center, Greenbelt, MD, USA\\
$^5$ University if California at Los Angeles\\
$^6$Southwest Research Institute, San Antonio, TX, USA\\
$^7$University of California, Berkeley, CA, USA}
\date{\today}

\begin{abstract} 
Plasma turbulence is investigated using high-resolution 
ion velocity distributions measured by the Magnetospheric Multiscale Mission (MMS) 
in the Earth's magnetosheath. 
The particle distribution is highly structured, 
suggesting a cascade-like process in velocity space. 
This complex velocity space structure is investigated
using a three-dimensional Hermite transform
that reveals a power law distribution of moments. 
In analogy to hydrodynamics, a Kolmogorov approach 
leads directly to a range of predictions for this phase-space cascade. 
The scaling theory is in agreement with observations, suggesting a 
new path for the study of plasma turbulence in weakly collisional 
space and astrophysical plasmas.
\end{abstract}
\maketitle 

Turbulence in fluids is characterized by nonlinear interactions
that transfer energy 
from large to small scales, eventually producing heat.
For a collisional medium, 
whether an ordinary gas 
or a plasma, 
turbulence leads to complex real space structure, 
but the velocity space, constrained by collisions,
remains smooth and close to local thermodynamic equilibrium 
(as, e.g., in Chapman-Enskog theory \cite{Huang}.)
However, in a weakly collisional plasma,  
spatial fluctuations are accompanied by 
fluctuations in velocity space,
representing another essential facet of plasma dynamics. 
The characterization of the velocity space is challenging 
in computations and in experiments,
although Vlasov simulation has 
revealed complexity in the velocity space, often 
near coherent magnetic and flow structures 
\cite{ServidioEA12,GrecoEA12,TenBargeEA13-sheets}.
Here we make use of powerful new spacecraft observations in the 
terrestrial magnetosheath that reveal
this structure with sufficient accuracy to 
quantify the velocity cascade for the first
time in a space plasma. 

The observations reported here are 
enabled by 
the Magnetospheric Multiscale Mission (MMS), launched in 2015 
to explore magnetic reconnection.
The MMS/FPI instrument measures ion 
and electron velocity distributions 
(VDFs) at high time cadence,
and with high resolution in angle and energy.
High resolution magnetic field measurements
are available and
four-point observation 
is available for all instruments.
MMS provides characterization of plasma turbulence with 
unprecedented resolution and accuracy. 
The spacecraft orbit repeatedly 
crosses the Earth's magnetosheath,
enabling new and important characterizations of plasma dynamics
(see e.g.  \citet{BurchEA16}).
Here we focus on one traversal of the magnetosheath, and 
specifically on a quantitative description of the ion velocity space cascade.

\begin{figure}
\includegraphics[width=\columnwidth]{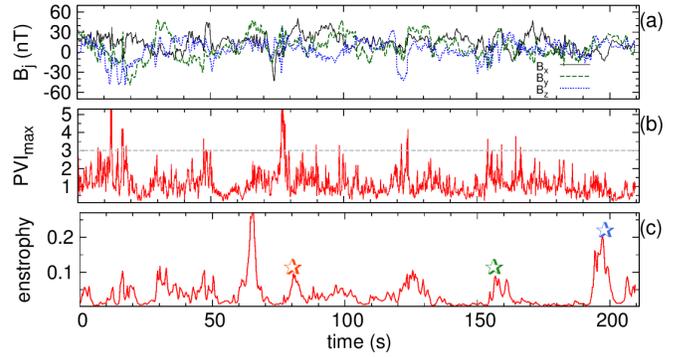}
\caption{Sample of MMS data, plotted vs. time (seconds), beginning at 
2016-01-11, 00:57:04 
in the magnetosheath. (a) Magnetic field components; 
(b) PVI$_{max}$, calculated as the maximum of the magnetic, 
velocity and density intermittent time series 
(horizontal line represents a typical threshold); 
(c) Enstrophy. Stars indicate selected times 
at which we show ion VDFs in Fig. 2.
Analyses of VDFs will refer to this time axis. }
\label{fig:mmsF}
\end{figure}

{\it Magnetosheath data sample.}
The analysis below employs 
data from 
the period 2016-01-11, 00:57:04 to 2016-01-11, 01:00:33, 
about five hours after 
an outbound magnetosheath crossing, 
and four hours before the next inbound crossing. 
Apogee is $\approx$12 $Re$ at 02:16:54.
The spacecraft, separated by $\sim 40$km
($\sim \frac12$ ion gyroradius) 
are downstream of the quasi-parallel bow shock, 
and the interplanetary magnetic field 
is nearly radial. In these conditions, fully developed 
upstream turbulence readily convects into the magnetosheath. 
The selected interval contains fine scale activity including 
sub-proton scale current sheets, as previously described in some 
detail by \citet{ChasapisEA17}. 
The magnetic field in this
period is very turbulent and structured, 
as reported in Figure \ref{fig:mmsF}-(a). 
In this fairly typical magnetosheath interval, 
the ratio $\delta b/B_0\sim 1.5$ ($rms$ of the fluctuations/mean field),  
indicating near-isotropic turbulence, while the plasma beta $\beta \sim 7$. 
The magnetic field power spectrum (not shown) 
manifests a clear Kolmogorov scaling
on inferred 
wavenumber $k \approx f/(2 \pi V)$ (bulk flow speed $V$), 
with $k^{-5/3}$ slope at larger scale, 
and a break-point near $f= 0.8$ Hz,
giving way to a steeper $k^{-8/3}$
spectrum at higher frequencies, 
a pattern also commonly observed in the 
solar wind plasmas \citep{SahraouiEA09}.

Coherent structures 
are present in the analyzed interval and are of 
significance.
Previous works \citep{ServidioEA14,ChasapisEA17} 
show that  
kinetic processes 
such as heating are connected 
to sharp gradients of magnetic, 
velocity and density fields. 
In simulations, such coherent structures are 
correlated with strong
distortions of the proton VDFs
\citep{ServidioEA12,GrecoEA12,ServidioEA15}.
To select structures, 
here we compute a multiple-data stream
variation of the 
Partial Variance of Increments method, 
defining 
$\text{PVI}_{max}(t)=\text{MAX}\left\{\text{PVI}_n(t), \text{PVI}_u(t), \text{PVI}_b(t)\right\}$, 
where, for each field $g$ (density $n$, velocity $\bm u$, magnetic field $\bm b$),
the value of PVI is computed in  the standard way as 
PVI$_g(t) = |\Delta {\bm g}|/{\sqrt{  \left\langle|{\Delta {\bm g}}|^2\right\rangle}  }$. 
In this implementation \citep{GrecoEA09}, PVI 
is based on increments $\Delta g$ evaluated at $0.3$ s lag and a time
average $\langle \dots\rangle$ computed over the full sample interval. 
This multiple-data PVI (see Fig. \ref{fig:mmsF}-(b))
is sensitive to magnetic  
as well as vortex and shock-like structures.

The remainder of our
analysis concentrates 
on the ion velocity distribution functions (VDFs) $f({\bm v}, t)$. 
The resolution of the original FPI 
ion VDF 
dataset is $\Delta t=150$ ms, and the total 
burst interval duration is $\sim210$ seconds, 
with data accumulated at 
each of the MMS spacecraft MMS$_i$, with $i=1, .., 4$. 
These VDFs in spherical geometry, 
$f(v, \phi, \theta, t)$, are 
collected in the spacecraft frame, 
with very high precision and angular 
resolution. Here $\phi$ 
is the azimuthal angle ($0<\phi<2\pi$), $\theta$ the angle with 
the $z$ (spin) axis ($0<\theta<\pi$), and $40<v<2400$ km/s. 
The FPI instrument achieves its highest time resolution 
by sampling 32 energy channels at each measurement time,
and an 
interleaved set of 32 energy channels 
at the following time. Averaging over pairs of 
data samples blurs the time
by $150$ ms, but doubles energy resolution. 
We use the averaged, higher energy-resolution   
data in the analysis below, with
$N=$64 log-spaced energy channels, 
$N_\phi=32$ and $N_{\theta}=16$ equally sampled angular channels. 
Averaging in this way also reduces velocity 
space data gaps. 

{\it Hermite analysis method.}
In order to characterize $f({\bm v}, t)$, we employ 
a 3D Hermite transform representation, 
a method well-suited for analytical and numerical study of plasmas \citep{Grad49,ArmstrongMontgomery67,SchumerHolloway98,ParkerDellar15}. 
The ``physicists'' Hermite  polynomials are a 
classical sequence, 
defined as $H_m(v)=(-1)^m e^{v^2}\fr{d^m}{d v^m}e^{-v^2}$, 
orthogonal in a Hilbert space where the metric is defined by the Maxwellian 
weight function $e^{-v^2}$. The one-dimensional basis functions 
are 
\beq
\psi_m(v)=\frac{ H_m\!\!\left(\!\fr{v-u}{v_{th}}\!\right) }{\sqrt{2^m m! \sqrt{\pi} v_{th} }} e^{ - \frac{(v-u)^2}{2 v_{th}^2}}, 
\lab{eq:psim}
\eeq
where $u$ and $v_{th}$ are the bulk velocity and the thermal speed, 
respectively, and $m\geq 0$ is an integer. 

The eigenfunctions in Eq.~(\ref{eq:psim}) obey 
the orthogonality condition 
$\int_{-\infty}^{\infty} \psi_m(v)\psi_l(v) d v = \delta_{m l}$. 
Using this basis, one can obtain a 3D decomposition of the distribution function
\beq
f({\bm v}) = \sum_{\bm m} f_{\bm m} \psi_{\bm m}( {\bm v} ), 
\lab{eq:decomp}
\eeq
where the 3D eigenfunctions are 
$\psi_{\bm m}( {\bm v} )=\psi(m_x, v_x)\psi(m_y, v_y)\psi(m_z, v_z)$, 
and the Hermite coefficients are 
\beq
f_{\bm m} = \int_{-\infty}^{\infty} f({\bm v}) \psi_{\bm m}( {\bm v} ) d^3 v.
\lab{eq:proj3d}
\eeq
Note that, in the case of a Maxwellian, $f(v)=M(v)=e^{-v^2/2}$ and the 
first coefficient $f_{\bm 0} = \fr{n}{\left[ 2 v_{th}  \sqrt{\pi}\right]^{3/2}}$. 
This simple case gives a deep meaning to the Hermite projection in plasmas, 
namely that each Hermite index $m$ roughly corresponds to an order of the plasma moments: 
the $m=1$ coefficient corresponds to bulk flow fluctuations;
$m=2$ corresponds to temperature deformations;
$m=3$ to heat flux perturbations, and so on. 
This suggests that highly deformed VDFs would produce a distribution of modes.

{\it MMS analysis.}
We now perform a Hermite analysis of the MMS data. 
Required projections are based on a 
3D non-uniform grid in each direction based on  
the zeroes $v_j$ of the order $N_v+1$ Hermite polynomial, 
[$H_{N_v+1}(v_j) = 0; j = 0, N_v+1$], thus 
defining a quadrature. 
For these results, we choose $N_v\equiv N_{v_x}=N_{v_y}=N_{v_z}=100$.
This spans a velocity space, centered at zero speed, 
defined by values $v_j$ as
$v_0 = - V_{max}, ..., v_{N_v/2}=0, ..., v_{N_v}=V_{max}$ \citep{Zhaohua14}.
The velocity is normalized in terms of the local thermal velocity $v_{th}$, 
the density is normalized such that $n=1$, and the local fluid velocity ${\bm u}=0$ 
is built into the representation as described above (velocity is measured relative to 
the bulk fluid frame). 
Values of 
$f({\bm v})$ are transformed from the
native (MMS) spherical representation
to the non-uniform (Cartesian) grid, 
using a 2$^{nd}$ order interpolation method, 
weighting with volumes 
$V = \int_{v_1}^{v_2} \int_{\theta_1}^{\theta_2} \int_{\phi_1}^{\phi_2} v^2 sin\theta~dv d\theta d\phi$
within each angular sector of the MMS data grid. 
This procedure produces a 
normalized  VDF on a new ``Hermite grid'', $f(v_x, v_y, v_z)$, 
where velocities are in units of local thermal speed, ${\bm u}=0$, 
and unit density. 
The occurrence of missing data points is 
reduced by averaging $f({\bm v})$ over the four MMS satellites. 
When the spacecraft are closely separated, 
as in the present case, 
there are no remarkable differences among the four MMS$_i$ datasets, 
and therefore this averaging does not change
the following results. 

\begin{figure}
\includegraphics[width=0.49\textwidth]{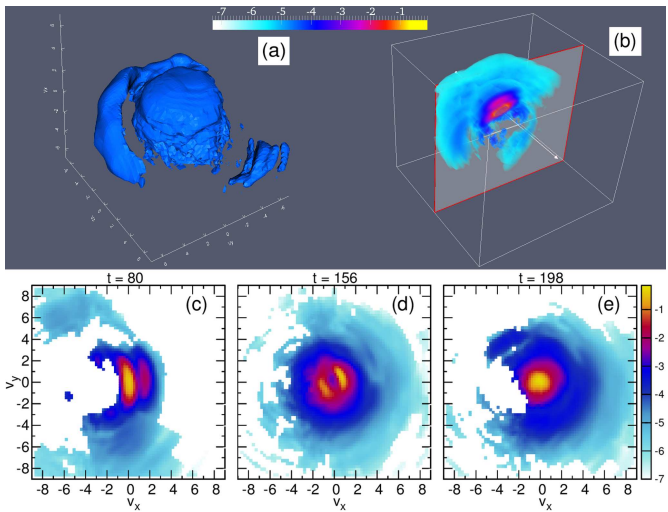}
\caption{(a) Ion velocity distribution function, obtained from the MMS mission interpolating the 
function over a Hermite grid (data from $t=80$ s in Fig.\ref{fig:mmsF}) 
and averaging over the 4 satellites. (b) 2D cut in the $v_x, v_y$ plane, with 3D shaded contours. Panels (c), (d) and (e) represents slices of the VDF 
at different times, highlighted with stars in Fig. \ref{fig:mmsF}-(c).}
\label{fig:vdf}
\end{figure}

Following this procedure results in 
a three dimensional rendering of the 
interpolated VDF at a single time ($t\sim 80$ s in Fig. \ref{fig:mmsF}),
as reported in Fig. \ref{fig:vdf}.  
The distribution is highly non-Maxwellian, as 
the iso-surfaces reveal the presence of secondary beams, rings of particles, multiple-anisotropies, 
heat flux and so on. 
This pictorial representation already 
suggests a broad spectrum of Hermite moments. 
The same figure shows 2D cuts 
of the interpolated VDF, at several different times:
at $t=80$ [same as panel (a); 
at $t=156$; 
and at $t=198$ s.  
Fig. \ref{fig:mmsF} provides the context.
It is evident that there are strong non-Maxwellian deviations, 
with differences varying in time (and due to the high speed flow in the 
magnetosheath, varying in spatial position). 
These deformations can be initiated by 
different local processes \citep{GrecoEA12,HowesEA17}.

In order to quantify deviations from fluids, we compute 
the mean square departure from Maxwellianity, 
which can also be described as 
the second Casimir invariant of the VDF, 
or, in analogy to the mean square vorticity in hydrodynamics \citep{Knorr77},
an enstrophy. 
We define 
the local deviation from the associated Maxwellian 
$\delta f = f(v) - M(v)$, that is 
equivalent to subtracting  $f_0$ from the Hermite series. 
Using this projection, indeed, the Parseval 
theorem gives the enstrophy
\beq
\Omega(t) \equiv \int_{-\infty}^{\infty} \delta f^2({\bm v}, t ) d^3 v = \sum_{\bm m>0} \left[f_{\bm m}(t)\right]^2.
\lab{eq:parce3}
\eeq
This quantity is zero for a pure Maxwellian, and may be compared with 
other parameters adopted as measures of non-Maxwellianity in plasma 
turbulence studies \citep{GrecoEA12}. It is also related 
to what is designated 
the ```free energy'' in certain reduced perturbative 
treatments of kinetic plasma 
(e.g., \cite{SchekochihinEA08}), except that in the 
present case no perturbation is implied.
The plasma enstrophy as a function of time 
is reported in Fig. \ref{fig:mmsF}-(c).  
Its behavior is quite bursty, and is 
qualitatively connected to the spatial intermittency of the system. 
For example, peaks of enstrophy frequently correspond to regions where 
PVI$_{max}$ is high, or nearby these maxima. 
This is consistent with previous studies \citep{GrecoEA12,ParasharMatthaeus16} that 
found anisotropic or non-Maxwellian features in the vicinity of 
magnetic coherent structures such as current sheets. 
The distributions shown above in Fig. \ref{fig:vdf} 
correspond to the times of local peaks of $\Omega(t)$ seen in Fig. \ref{fig:mmsF}-(c). 

\begin{figure}
\includegraphics[width=0.49\textwidth]{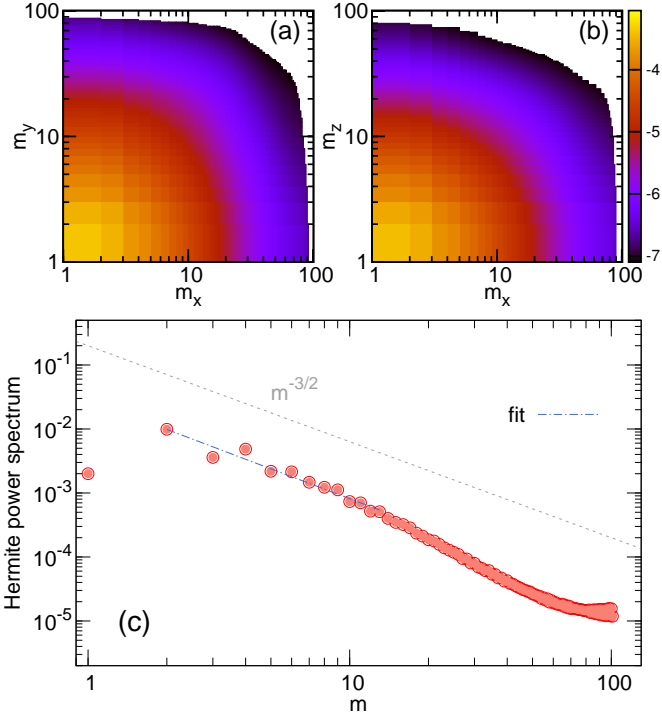}
\caption{(a) and (b): 2D reduced Hermite spectrum, indicating spectral isotropy 
in these velocity space planes. 
(b) Power spectrum of the Hermite modes for the MMS dataset.
The best fit to a power law $m^{-\alpha}$ gives $\alpha\sim1.5$, with an error of $\sim7\%$.}
\label{fig:mmslaw}
\end{figure}

Following  Eq.~(\ref{eq:proj3d}), we compute 
the modal 3D Hermite  
spectrum $f_{{\bm m}}^2(t)$. 
For an ensemble average 
description of the entire sample, 
our method
averages the multidimensional Hermite spectrum over time,  
indicating this as 
$E(m_x, m_y, m_z)=\langle f_{{\bm m}}^2(t)\rangle_T$.  
The 3D modal spectrum (as in Fourier analysis)
permits examination of the 
full 3D structure of the 
spectral distribution. Given the
great volume of data, 
it may be reduced or sampled 
to attain more compact representations. 
To this end, 
we compute the reduced 2D spectra as 
$E(m_x, m_y) = \sum_{m_z} E(m_x, m_y, m_z)$,
and analogously $E(m_x, m_z)$ and $E(m_y, m_z)$. 
Figure \ref{fig:mmslaw} shows two of these
reduced spectra. Within their respective planes, 
these spectra are quite isotropic, indicating the absence 
of strong magnetization and/or other privileged directions
in this stream. (Recall that plasma $\beta \sim$ 7.)

Based on the 2D spectra,
a reasonable 
way to characterize 
the velocity space fluctuations 
for this dataset 
is the isotropic velocity space spectrum. 
The isotropic (omni-directional) Hermite spectrum, 
in analogy to the 
classical spectral density in hydrodynamic turbulence, 
is computed by summing $m_x,m_y,m_z$ 
over concentric shells of thickness $\delta$ (here, unity) 
in the Hermite index space. 
That is, 
$P(m) = \sum_{m-\frac12 < |{\bf m'}| \leq  m+\frac12} 
E({\bf m}') $.  
The isotropic Hermite spectrum of magnetosheath 
turbulence is reported in Fig. \ref{fig:mmslaw}-(c). 
The velocity space distribution 
follows a power law behavior through at least the first ten moments, 
indicating the possibility of a phase-space turbulent cascade, 
as suggested in the literature  
\citep{SircombeEA06,SchekochihinEA08,Tatsuno09,SchekochihinEA16,KanekarEA15,ParkerDellar15}. 

{\it The model.} The analysis suggests a cascade in the plasma moments, analogous
to the classical real-space fluid cascade. 
To describe the inertial range of this 
cascade, $P(m)\sim m^{-\alpha}$, 
seen in Figure \ref{fig:mmslaw}-(c) at $m<15$, 
we develop a turbulence theory based on qualitative arguments, 
in style similar 
to the Kolmogorov phenomenology \citep{Kolmogorov41a}. 
The Boltzmann kinetic equation for weakly-collisional plasma reads as
\begin{eqnarray}
\frac{\de f}{\de t} + {\bm \nabla}\cdot{\left( {\bm v} f \right)} + \frac{e}{M_p}\left({\bm E} + {\bm v}\times{\bm B} \right)\cdot {\bm \nabla}_v f = C_\nu. 
\label{eq:bltz}
\end{eqnarray}
This equation 
couples with the Maxwell equations to determine 
the electric field $\bm E$ and magnetic field $\bm B$.
$M_p$ indicates the mass and $e$ the charge of ions. 
The r. h. s. of Eq.~(\ref{eq:bltz}) is a collision operator, 
which may have a complex form. 
We make use presently 
of two Hermite recursion relations:
\bea
v~\psi_m(v) = \sqrt{\frac{m}{2}} \psi_{m-1}(v) + \sqrt{\frac{m+1}{2}} \psi_{m+1}(v), 
\lab{eq:rec1}
\\
\fr{\de \psi_m(v)}{\de v} = \sqrt{\frac{m}{2}} \psi_{m-1}(v) - \sqrt{\frac{m+1}{2}} \psi_{m+1}(v).
\lab{eq:rec2}
\eea
Upon computing 
Hermite and Fourier transforms of Eq.(\ref{eq:bltz}), 
and 
using Eq.s~(\ref{eq:rec1})-(\ref{eq:rec2}), 
one arrives at an evolution equation for the coefficients 
$\frac{\partial f_{\bm m}({\bm k}, t)}{\partial t}$, 
in a 7D space (3D Fourier space, 3D Hermite space, and time).
Because of the complexity of this equation, we will proceed with some ansatz, 
an approach familiar in Navier-Stokes turbulence. 
First, we neglect the collisions, 
which are likely to be confined 
to very high Hermite modes $m$'s. Second, we envision 
three (asymptotic) regimes, assuming locality in scale,
\begin{equation}
  \frac{ \de f_m(k, t)}{ \de t} \sim \left \{
  \begin{aligned}
    &k\sqrt{m}v_{th}~\mathcal{S}_m\!\!\left\{f_m(k, t)\right\} && (a), \\
    &\frac{e E(k)}{M_p v_{th}}\sqrt{m}~\mathcal{D}_m\!\!\left\{f_m(k, t)\right\} && (b), \\
    &\frac{e B(k)}{M_p} m~\mathcal{A}_m\!\!\left\{f_m(k, t)\right\} && (c).
  \end{aligned} \right.
  \label{eq:approX}
\end{equation} 
These regimes are obtained immediately from 1D-1V models 
using  Eq.s~(\ref{eq:rec1})-(\ref{eq:rec2}), assuming that
[case (a)] the dominant term is either due to bulk and thermal fluctuations;
or [case (b)],
that the main fluctuations are due to the electric field; or [case (c)],
that the dynamics is governed by the magnetic field. 
Here 
$E(k)$ and $B(k)$ represent the electric and magnetic spectra. 
In Eq. (\ref{eq:approX}), $\mathcal{S}_m$ is a linear operator, similar to a splitting operator that propagates information into $m+1$ and $m-1$;
$\mathcal{D}_m$ is an operator that 
resembles derivatives in the $m$-space \citep{SchekochihinEA16}; 
and $\mathcal{A}_m$ is a higher order operator, 
combinations of the previous two. 
Note that Eq.~(\ref{eq:approX})-(c) also introduces anisotropy in the $m$-space, 
which we will  discuss in future works.  
In all three limits we expect 
redistribution of fluctuations in $m$-space, 
though a cascade/diffusion-like process. 
Following the Kolmogorov intuition, 
using now Eq. (\ref{eq:approX}), we extract 3 characteristic 
times that depend on both $k$ and $m$:
\bea
\lab{eq:taukmV}
\tau_{v}(k, m)&=&\frac{1}{v_{th} k \sqrt{m}}\sim m^{-1/2} \\
\lab{eq:taukmE}
\tau_E(k, m)&=&\frac{M_p v_{th}}{e E(k)\sqrt{m}}\sim m^{-1/2} \\
\lab{eq:taukmB}
\tau_B(k, m)&=&\frac{M_p }{e B(k) m}\sim m^{-1}.
\eea
These times measure in some sense the relative 
intensity of the corresponding terms 
in the dynamical system Eq.~(\ref{eq:bltz}).

At this stage, we 
adopt the hypothesis of an enstrophy cascade in 
the velocity space. 
We integrate over volume, and therefore over $k$,  
to find a scaling law in $m$, based on the
idea that a velocity cascade proceeds conserving the 
quadratic ``rugged'' invariant $\Omega$, 
defined in Eq.(\ref{eq:parce3}). 
The first hypothesis 
is that there is a net constant flux in the $m$-space, namely
\beq
\epsilon = \frac{f_m^2}{\tau_m} = const.
\lab{eq:flux}
\eeq
where $\tau_m$ is the spectral transfer time for the enstrophy. 
The second hypothesis concerns choice of the 
characteristic time of this cascade, the simplest options being to 
chose among (\ref{eq:taukmV})-(\ref{eq:taukmB}).
Third (and last), from simple dimensional arguments, 
\bea
\nn
&\Omega = \langle \int \delta f^2 d^3 v \rangle_x= \sum_{m>0} f_m^2 =  \int P(m) dm ,& \\
&\rightarrow P(m)\sim f_m^2 m^{-1},& 
\lab{eq:omegaa}
\eea
where we defined $\langle ...\rangle_x$ as the physical space volume average. Using Eq.s (\ref{eq:flux}), (\ref{eq:omegaa}) and a characteristic time such as (\ref{eq:taukmV}) or (\ref{eq:taukmE}), one gets
\beq
P(m) \sim  m^{-3/2}.
\lab{eq:law1}
\eeq
Analogously, using Eq.s~(\ref{eq:flux}) and (\ref{eq:omegaa}), coupled with (\ref{eq:taukmB}) one obtains
\beq
P(m)\sim m^{-2}.
\lab{eq:law2}
\eeq
The first law in Eq. (\ref{eq:law1}) should be valid in a thermal 
and/or electric-dominated regime, while the last prediction 
is more adequate for a highly magnetized (anisotropic) plasma. 
For the present observation, 
we fit an inertial range powerlaw to the velocity-space cascade, 
as shown in Figure \ref{fig:mmslaw}-(c), 
obtaining $P(m) \sim  m^{-\alpha}$, with $\alpha=1.5\pm 0.1$, in agreement with Eq.~(\ref{eq:law1}). 

To summarize and conclude, we have carried
out an analysis of MMS ion VDF data to visualize 
and describe the ion distribution function in this 
low-collisonality space plasma with  
unprecedented temporal and velocity-scale resolution.
We observe 
here in spacecraft data the same kind of 
fine scale velocity structure reported frequently in 
Vlasov simulations \cite{ServidioEA15,HowesEA17}.
This motivates a further analysis of the velocity 
space structure in terms of a Hermite spectral analysis,
which has the physically interesting heuristic interpretation 
as a moment heirarchy. The power law that emerges 
in moments (Hermite indices) suggests a velocity space 
cascade. 
We pursue this in a very preliminary way, in analogy to
classical hydrodynamics cascade. 
One first identifies  
 a conserved flux across scale 
-- here the velocity space enstrophy (or, free energy) -- and 
the associated dynamical time scales. 
From this emerges the possibility of spectral slopes 
between -2 and -3/2. 
Other possibilities may 
exist for other physical regimes in which different 
time scales become available.
For the MMS magnetosheath interval analyzed here, 
the -3/2 slope seems 
to be clearly favored, suggesting   
that the velocity space 
cascade for this interval 
is governed by thermal and/or electric effects 
\citep{ValentiniVeltri09}.

This observation and analysis is preliminary, 
being based on a single set of high resolution observations, and so 
we must eschew any assignment of universality. 
However, enabled by significant advances in diagnostics such 
as those offered by MMS, this approach to understanding 
velocity space structure may prove to be fruitful for 
further studies in turbulent plasmas, in varying conditions.

\begin{acknowledgments}
This research was partially
supported by AGS-1460130 (SHINE),
NASA grants NNX14AI63G (Heliophysics Grand Challenge Theory),
the Solar Probe Plus science team (ISOIS/SWRI subcontract No.
D99031L), and by the MMS Theory and Modeling team, NNX14AC39G.
F. V. is supported by Agenzia Spaziale Italiana under 
contract ASI-INAF 2015-039-R.O.
DP and SS acknowledge support from the 
Faculty of the European Space Astronomy Centre (ESAC).

\end{acknowledgments}


\end{document}